\documentclass[twocolumn]{openjournal}  

\usepackage{graphicx}
\usepackage{txfonts}
\shorttitle{SiO masers and 7 mm Continuum in Mira and R Aqr}
\shortauthors{Cotton et al.}
\begin{document} 

   \title{SiO masers and 7 mm continuum emission in binary AGB stars
     Mira and R Aqr.}

   \author{W.~D.~Cotton}
   \affiliation{National Radio Astronomy Observatory, 520 Edgemont Road, Charlottesville, VA 22903, USA}
   \author{E.~Humphreys, M.~Wittkowski}
   \affiliation{
     European Southern Observatory, Karl-Schwarzschild-Str. 2, 85748 Garching bei Munchen, Germany}
   \author{A.~Baudry}
   \affiliation{
     Laboratoire d'astrophysique de Bordeaux, Université de Bordeaux, 33615 Pessac, France }
    \author{A.~M.~S.~Richards}
    \affiliation{JBCA, School of Physics and Astronomy, University of Manchester, M13 9PL, UK }
    \author{W.~Vlemmings,  T.~Khouri}
    \affiliation{ Department of Space, Earth and Environment, Chalmers University of Technology, 
      Onsala Space Observatory, 439 92 Onsala, Sweden}
    \author{S.~Etoka}
    \affiliation{JBCA, School of Physics and Astronomy, University of Manchester, M13 9PL, UK }



 
\begin{abstract}
   {Interactions between AGB stars and a secondary in a close orbit
     are one possible explanation of why some AGB stars develop into
     aspherical planetary nebulae.}
   {This study uses millimeter observations of the continuum and SiO
     maser emission in several symbiotic Miras looking for evidence of
   an interaction between the two stars.}
   {New JVLA observations at $\sim$45 mas resolution are analyzed, imaging
     continuum and SiO masers.}
   {Two of the three targets were detected and accurately registered
     continuum and line images were derived.}
   {No clear evidence of an interaction was found between components B
     and A in Mira.   
     R Aqr has a well known jet arising from the secondary star.  
     The jet may be disturbing the circumstellar envelop of the AGB
     star or possibly just nearly aligned with it. }
\end{abstract}

   \keywords{Stars: AGB and post-AGB -- Stars:binaries:symbiotic --
     Radio lines:stars -- Radio continuum:stars}
%

\section{Introduction}
A substantial fraction of main sequence stars, up to 50\%, are known
to exist in systems with multiple components.
The importance of close binaries on stellar evolution has been
recognized lately, but their effects on the close circumstellar
environment is under active study. 
The influence of a companion on the region near to the surface of
Asymptotic Giant Branch (AGB) stars may play a pivotal role in the
formation of aspherical Planetary Nebulae (PNe).  
Observations of molecules in the winds of a sample of AGB stars
\citep{Ramstedt2014,Ramstedt2017,Ramstedt2018,Doan2017,Decin2020} show
a variety of non spherical shapes which are 
attributed to the interactions with companion stars.
Similar shaping might affect the episodes of mass loss leading to the
formation of PNe.

SiO masers are usually found within the pulsating atmospheres of AGB
stars at about 2 stellar radii, at about the distance at which
silicate and other dust forms in the outflows 
\citep{Karovicova2013,Gobrecht2016,Hoefner2019}.
These masers generally appear in the form of a ring centered on the
photosphere and reveal motions in the inner envelope
\citep{Boboltz1997,Hollis1997b,Cotton2004,Cotton2006,Ragland2008}.
Single dish monitoring of semi-regular and Mira AGB stars by
\cite{Gomez-Garrido2020} show large and rapid variations of the SiO
masers in the semi-regular giants RX Boo and RT Vir and slower
variations in several Miras.

One of the current unknowns in stellar evolution is the shaping
mechanism from asymptotic giant branch (AGB) stars to planetary
nebulae. 
On the one hand, both recent optical spectro--polarimetry
and radio/submm maser observations indicate unexpectedly high magnetic
fields near or at the stellar surface \citep{Lebre2014,Vlemmings2018}
and references therein. 
On the other hand, ALMA observations indicate the presence of
companions around stars that were previously believed to be isolated
e.g. the iconic ALMA observations of R Sculptoris;
\citep{Maercker2012}.  
Whether binarity or magnetic fields dominate the evolutionary shaping
mechanism is a matter of current debate. 

One of the best places to study how the out-flowing material is shaped
is at, or near, the stellar surface. 
Long baseline ALMA observations of the surfaces of AGB stars Mira and
W Hya reveal hot-spots that likely influence and cause asymmetry in the
near--circumstellar environment \citep{Vlemmings2015,Vlemmings2017}.
Towards binary system Mira AB, we have also recently discovered signs
of the influence of the white dwarf Mira B at only a few stellar radii
from AGB star Mira A (Humphreys et al., in prep). 
In this case, using ALMA long baseline data we found that SiO masers
trace a portion of a bubble wall formed by the interacting Mira AB
winds, also seen as a spiral plume in SiO thermal gas.
In this paper, we present observations of continuum and SiO maser
lines using the JVLA of Mira and two other AGB stars with known
companions, W Aql and R Aqr. 
Our aims are to study the effect of the companions on the
observed maser spot distribution. 

\cite{Cotton2004,Cotton2006} 
give a sequence of VLBA images of the SiO masers in the atmosphere of Mira.
These appear as partial rings with diameters in the range 60 to 75
mas.
The diameter of the $\nu$=1,J=1-0 maser ring tends to be slightly larger
than that of the $\nu$=2,J=1-0 masers.
\cite{Mennesson2002} give a uniform disk diameter of the photosphere in
K band of 28.79 $\pm$ 0.10 mas.
\cite{Wittkowski2016} give a diameter of 28.5 $\pm$ 1.5 mas at 2.25
$\mu$m.
\cite{Perrin2020} give the photospheric diameter in H band of 21.1 mas.
The distance to Mira is given by \cite{Hannif1995} as 110 $\pm$ 9
pc. at which distance 1'' corresponds to 110 AU.
The Hipparcos distance \citep{vanLeeuwen2007} is 92$\pm$11 pc.

W Aquilae (W Aql)  is a binary S--type AGB star sometimes showing SiO
maser emission. 
SiO maser emission observed towards W Aql \citep{Nakashima2007} 
disappeared for a time around 2011 \citep{Ramstedt2012}, which
could be due to disruption by an additional companion (the known
companion lies at about 0.46 arcseconds from the S--type AGB star,
about 180 AU at a distance to W Aql of 395 pc). 
Peak SiO maser emission can reach 21 Jy (single--dish;
\cite{Ramstedt2012}).  

R Aquarii (R Aqr) is a  symbiotic binary and a well--known host of
millimeter and submillimeter SiO masers \citep{Boboltz1997,
Hollis1997b,Gray1998,Cotton2004,Cotton2006,Ragland2008}.
\cite{Cotton2004,Cotton2006} give a sequence of VLBA images
of the SiO masers in the atmosphere of R Aqr with ring diameters of 31
to 33 mas.
Infrared measurements of the photospheric size show a range of values
between 11.2 and 18.4 mas \citep{Tuthill2000,Mennesson2002,Ragland2008,Wittkowski2016}.


The maser emission is variable but peak values in single--dish
observations lie in the range up to 300 Jy. 
R Aqr consists of a Mira variable and an accreting, hot companion
with a remarkable jet outflow. 
HST observations of the jet shows considerable change over a several
year timescale \citep{Hollis1997a}. 
In 2014 the stars were separated by 45 mas \citep{Schmid2017} or 9.8 AU
at the distance to R Aqr of 218 pc \citep{Min2014}.
ALMA observations of \cite{Bujarrabal2018} detect both stars with a
bridge of material joining them, likely material flowing from the AGB
primary to the accretion disk around the WD secondary.
\cite{Cotton2004} show an unusual rotating ring of masers in January
2001. 
This ring was interpreted as arising in a rotating equatorial disk of
material.

\section{Observations}
The observations were made on the JVLA near 7 mm wavelength during 2019
August 12 from 06:00:00 UT to 13:00:00 UT under project code 19A-220
in the most extended (``A'') configuration. 
The target stars (astrometric/bandpass calibrators) were W Aquilae
(J1939-1525), R Aquarii (J2348-1631) and Mira, AKA omicron Ceti,
(J0217+0144). 
The pulsation phase for W Aquilae was 0.27, for R Aquarii was 0.21 and
0.79 for Mira.
The photometric calibrator was 3C48.

The observing sequence consisted of cycling between one of the targets (7 min)
and its calibrator (2 min) for an hour and then proceeding to another
target.
Reference pointing was used with a pointing measurement made on the
calibrator at the beginning of each sequence.
The integration time was 2 seconds and only the parallel hand data
(RR,LL) were recorded.

The masers observed were the SiO $\nu$=2,J=1-0 and
$\nu$=1,J=1-0  transitions at 42.820587 and 43.122027 GHz
respectively.
Each transition was covered by 256 channels of 62.5 kHz
bandwidth($\sim$0.9 km/s after Hanning). 
The center LSR velocity of the observation was 47 km/sec for Mira and
-22 km/sec for R Aqr and W Aql for which the true systemic velocity is
approximately -26 and -18 km/sec respectively.
The continuum observations consisted of 58 spectral windows in the
range of 41.0 to 48.7 GHz using 7424 $\times$ 1 MHz channels; spectral
regions containing the masers were not included.

\section{Calibration}
The conditions of these observations, Summertime, high frequency and
high resolution render the proper calibration of the data more
difficult. 
In order to allow accurate comparison of the continuum and line images
they must be aligned both photometrically and especially
astrometrically.
The relatively poor phase coherence of the data make this more difficult.
Photometric calibration is complicated by the relatively poor
sensitivity and strong resolution of the photometric calibrator by the
$\sim$45 mas resolution of the given data.
Fortunately, the masers in two of the targets are quite
strong and allow relative phase calibration following the general
approach of \cite{RM90,RM97,RM07}, 
adapted to the wide-band nature of the present data.

Data were translated from the archive format to AIPS format and were
processed in the Obit package \citep{OBIT}.
Hanning smoothing was used on the line data to
suppress the artifacts resulting from the interaction of the strong,
narrow maser signals and the finite delay range of the correlator.

\subsection{Amplitude \label{ampcal}}
The traditional amplitude calibration scheme for connected element
interferometers is to compare the gain solutions of the astrometric
and photometric calibrators to infer the flux density of the
astrometric calibrator.
This works well in the high signal-to-noise regime with well modeled
and only marginally resolved calibrators but can produce poor results
with low SNR data with strongly resolved photometric calibrators
as is the case here.
All of the centimeter wavelength stable photometric calibration
sources, including 3C48 used in this work, are strongly resolved at
the $\sim$45 mas resolution of this data.

We have adopted an alternate approach of using an initial, assumed
flux density for the astrometric calibrator and apply this calibration
to the photometric calibrator.
The ratio of the apparent integrated flux density of the derived image
of the photometric calibrator to its true flux density gives the
factor needed to correct the assumed flux density of the astrometric
calibrator. 
Astrometric calibrators are generally physically very small sources,
hence variable, but are well modeled by a point at the current
resolution. 
This approach has the advantage that a super accurate structural model
of the photometric calibrator is not needed but only an accurate total
flux density.
The adopted flux density of 3C48 at 40.98 GHz was 0.611 Jy and at 48.6
GHz was 0.506 Jy \citep{Perley}.
The systematic error in the amplitude scale is estimated to be 20\%.

\subsection{Phase}
The phase calibration approach of \cite{RM90,RM97} is to
use the phase calibration derived from a self calibration of a strong,
simple maser to calibrate the phase of the continuum data.
This can work well in the present case of AGB stars which may have very
strong, non-thermal masers but very weak thermal emission from the
photosphere. 

The original scheme of \cite{RM90,RM97}  was appropriate for
a narrow band system for which a standard group delay calibration was
adequate and did not need to be determined from the data-set in
question.
The continuum bandwidth of the current data is sufficiently wide that
corrections to the group delay need to be determined  and applied to
the data. 

The ``phase'' corrections derived from the self calibration of a maser
are largely due to tropospheric refraction, hence are narrow band
samples of a group delay function and this needs to be taken into
account in transferring the maser calibration to the continuum data
which extends for several GHz from the frequency of the maser.
Furthermore, the signal from a given maser spot is very narrow band
and inadequate to measure the relevant group delay.
The calibration method used is similar to that described by
\cite{Matthews2015,Matthews2018}.

We have adopted the following scheme.
\begin{enumerate}
\item {\bf System temperature calibration}.
Amplitude corrections were determined using program SYGain from the
on--line calibration measurements. 
\item {\bf Group delay calibration}.
The continuum observations of the phase reference sources were used to
determine the group delay error (program Calib). 
These calibrations were transferred to the maser line data and used to
calibrate both data sets.
\item {\bf Bandpass calibration}.
The observations of the phase reference sources were used to derive a
bandpass response function for each of the line and continuum data for
each target star using program BPass.
\item {\bf External calibration}.
The amplitude calibration scheme described in Section \ref{ampcal} was
used to derive the flux densities of the astrometric calibrators.
These were then used to obtain complex gain corrections (program
Calib) which were applied to the data.
\item {\bf Self calibrate a  maser}.
A strong, simple maser appearing in a few line channels was imaged
by SCMap for each target using self calibration.
After several cycles of phase only calibration, an amplitude and phase
self calibration was done.
\item {\bf Transfer phase to continuum data}.
The time sequences of gain corrections from the maser self calibration
were transferred to the continuum data extrapolating phase in
frequency using frequency scaling. 
This is done with program SnCpy.
\item {\bf Doppler corrections}.
Program CVel was used to correct the spectroscopic data for the
Earth's motion.  
\item {\bf Image}.
The continuum data were imaged with no further calibration producing a
wide-band image using program MFImage.
Channels in which maser emission might appear were excluded from the
continuum imaging. 
A spectral cube was imaged (via Imager) from the maser data for each
transition using the calibration from the self calibration.
\end{enumerate}

Use of self calibration to set the coordinates of the derived
images results in a large uncertainty of the absolute positions.
However, the resulting images should have accurate relative positions
of both masers and continuum emission with respect to the reference
maser. 

\section{Results}
No maser or continuum emission was detected from W Aql but strong
masers were detected in Mira and R Aqr and were used to calibrate the
continuum. 
Frequency channels with maser emission are identified from long term, scalar
averaged, visibility spectra from individual baselines.  
The mean channel (0.9 km sec$^{-1}$ ) visibility amplitude on a given
baseline for W Aql is 18 mJy and 5 times this should be easily
detectable, a plausible upper limit to a maser peak is 90 mJy. 
Pimpanuwat et al. (in prep.) report that ALMA observations of the
ATOMIUM
\footnote{https://fys.kuleuven.be/ster/research-projects/aerosol/atomium/atomium}
project detected thermal but not maser emissions of several 1.2 mm SiO
J=5-4 and J=6-5 transitions in W Aql during June and July 2019.

Many maser features are well separated in velocity and are quite small;
centroids can be measured to a small fraction of the resolution.
The maser rings can therefore be defined with an accuracy much
better than the size of the stellar photosphere.
The noise contribution to the position uncertainty of an unresolved,
isolated component is $\sim \theta_{FWHM}/(2\times {\rm SNR})$ where
$\theta_{FWHM}$ is the full width at half maximum of the resolution and
${\rm SNR}$ is the signal-to-noise ratio.
Since the spatial and frequency resolution of this data is such that
masers overlap in space and velocity, a list of ``masers'' was derived
by fitting one or two Gaussians to the image in alternate channels in
each of the transitions.

\subsection{Mira}
\begin{figure}
  \includegraphics[width=2.7in,angle=-90]{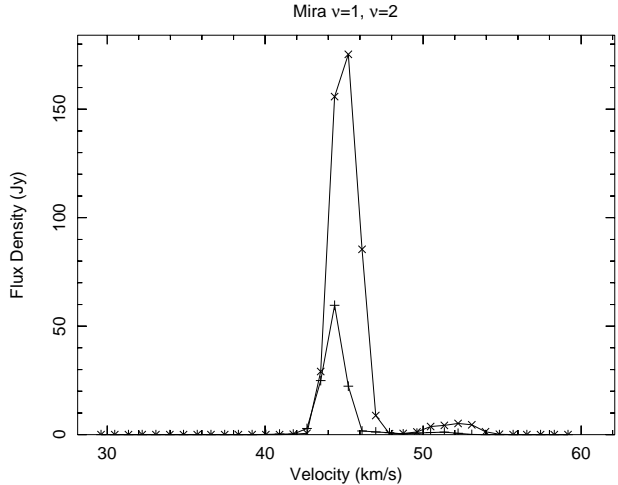}
  \caption{Spectrum of Mira SiO masers, ``+'' is $\nu$=1,J=1-0, and
    ``x'' is  $\nu$=2,J=1-0.}
  \label{FigMiraSpec}%
\end{figure}

The off--source channel noise in the maser cubes is 3.5 and 3.0 mJy
beam$^{-1}$ for $\nu$=1,J=1-0 and $\nu$=2,J=1-0 respectively and the
CLEAN restoring beams were  51 $\times$ 45 mas at position angle
41$^\circ$ and 50 $\times$ 40 mas at position angle -27$^\circ$.
The integrated spectra of the SiO masers in Mira are shown in Figure
\ref{FigMiraSpec}.

\begin{figure}
  \includegraphics[width=3in]{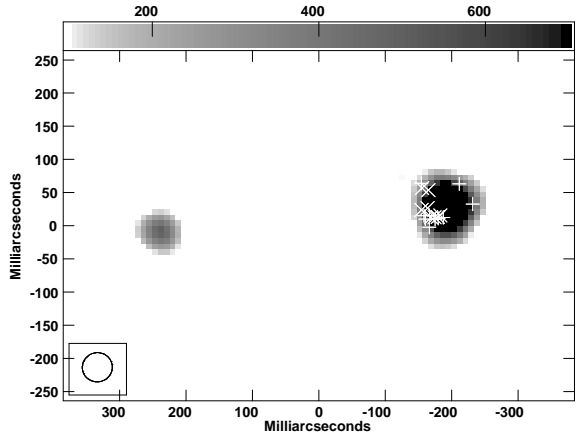}
  \caption{Continuum reverse gray-scale of Mira with ``+'' marking
    the locations of $\nu$=1,J=1-0 SiO masers and ``x'' for
    $\nu$=2,J=1-0 masers.
    A scalebar for the grayscale in $\mu$Jy Beam$^{-1}$ is given at the top.
    The resolution is shown in the box in the lower left corner.}
  \label{FigMira}%
\end{figure}

\begin{figure}
  \includegraphics[width=3.5in]{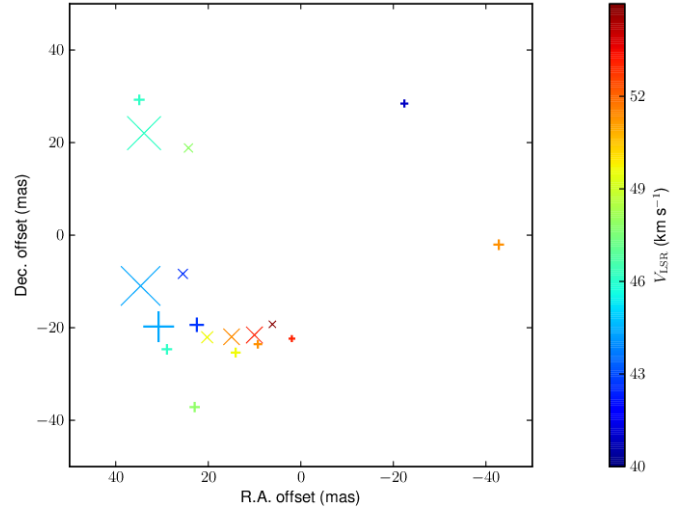}
  \caption{Color coded velocities of SiO masers surrounding Mira A
    with ``+'' marking  the locations of $\nu$=1,J=1-0 SiO masers and ``x'' for
    $\nu$=2,J=1-0 masers.
    Velocities are indicated by the color with a scale bar to the right.
    The size of each symbol is proportional to the $\sqrt{\quad}$ of the
    peak flux density.} 
  \label{FigMira_Vel}%
\end{figure}

The locations of the masers are plotted on the gray scale of the
continuum in Figure \ref{FigMira}.
The velocities and relative strengths of the masers are shown in
Figure \ref{FigMira_Vel}.
The SNR of the masers seen in Figure \ref{FigMiraSpec} exceeds 100 so
the uncertainty due to noise of the maser locations is less than 0.25
mas, much less than the maser ring size of 60 - 70 mas
\citep{Cotton2004} or the photospheric size of 28.79 mas
\citep{Mennesson2002}. 
The velocity ranges of the masers in the two transitions isn't the
same but this is also the case at several epochs shown in
\cite{Cotton2006} and is common in the single dish spectra of
\cite{Gomez-Garrido2020}. 
The continuum of the secondary, Mira B, is clearly detected but no SiO
was detected except at the location of Mira A.
No continuum was detected away from the two stars.

The off-source RMS in the continuum image is 28 $\mu$Jy beam$^{-1}$,
and the CLEAN restoring beam is 41 $\times$ 39 mas at position angle
-79$^\circ$. 
The peak of Mira B is 434 $\pm$ 3 mas (53 $\pm$ 0.3 AU) from Mira A at
an orientation of 95$^\circ$ from north through east.
The peak flux density of Mira A is 2.7 $\pm$ 0.6 mJy. 
The total flux density of Mira A is 5.0 $\pm$ 1.0 mJy  
with a deconvolved Gaussian size (FWHM) of 39.4 $\pm$
1.5 mas $\times$ 32.9 $\pm$ 1.5 mas at position angle -56.1$^\circ$
$\pm$ 4$^\circ$.  

Mira A being only moderately resolved and its profile, hence
area, being not well constrained by the present data, two possible
extremes are a uniform disk and a Gaussian.
We directly use the  factor of 1.6 in \citep{Pearson1999} to convert the
Gaussian size to an equivalent uniform disk size. 
We thus get for this extreme possibility a uniform disk diameter
of 63.0$\times$52.6 mas.
The brightness temperature of Mira A calculating its area
assuming this uniform disk and a total flux density of 5.0 mJy is 1300 $\pm$
270 K.
The brightness temperature calculating its area assuming a 
Gaussian profile with the flux density of 5.0 mJy is 2320 $\pm$ 480K.
The true brightness temperature is likely between 1300 and 2300 K.
Our total flux density was obtained by an intergral over the
image pixels and its error is dominated by the 20\% estimated
uncertainty in the flux density scale. 
The errors on the Gaussian size include the effects of correlated
noise in the image \citep{Condon97}.
The true uncertainty in the brightness temperature is dominated by the
uncertainty in the effective area of the source.


The peak flux density of Mira B is 0.62 $\pm$ 0.14 mJy with an
integrated value of 0.64 $\pm$ 0.14 mJy.
This component is at most marginally resolved with an upper limit
of 20 mas.

\subsection{R Aquarii}
\begin{figure}
  \includegraphics[width=2.7in,angle=-90]{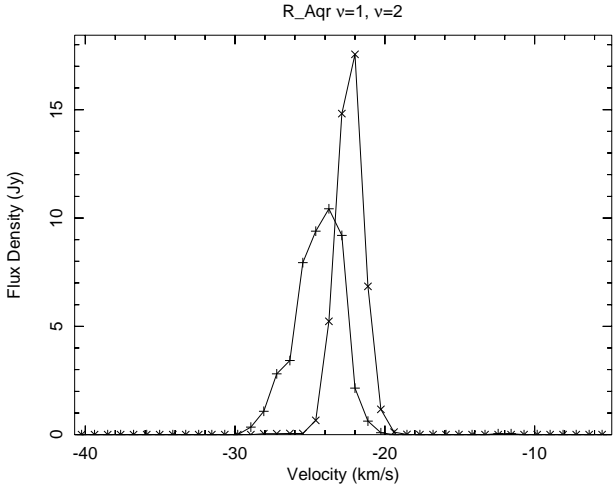}
  \caption{Spectrum of R Aquarii SiO masers, ``+'' is $\nu$=1,J=1-0, and
    ``x'' is  $\nu$=2,J=1-0.}
  \label{FigR_AqrSpec}%
\end{figure}

The off--source channel noise in the maser cubes is 4.4 and 3.9 mJy
beam$^{-1}$ for $\nu$=1,J=1-0 and $\nu$=2,J=1-0 respectively and the
CLEAN restoring beams were  59 $\times$ 41 mas at position angle
-4$^\circ$.
The integrated spectra of the SiO masers in R Aqr are shown in Figure
\ref{FigR_AqrSpec}.

\begin{figure}
  \includegraphics[width=3in]{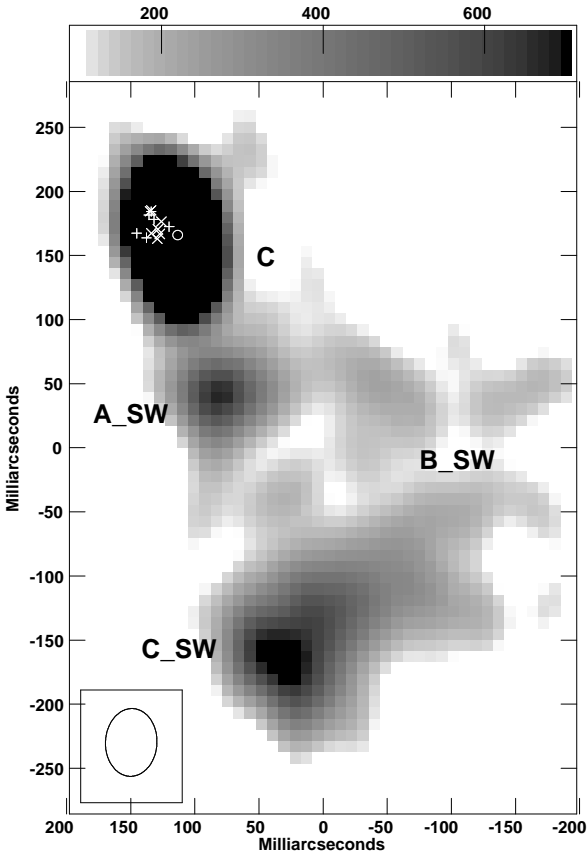}
  \caption{Continuum reverse grayscale of R Aqr with ``+'' marking
    the locations of $\nu$=1,J=1-0 SiO masers and ``x'' for
    $\nu$=2,J=1-0 masers.
    A scalebar for the grayscale in $\mu$Jy Beam$^{-1}$ is given at
    the top.
    Labels mark features approximately corresponding to those in
    \cite{Schmid2017}. 
    The circle shows the location of the secondary as estimated from the
    orbit of \cite{Bujarrabal2018}.
    The resolution is shown in the box in the lower left corner}
  \label{FigR_Aqr}%
\end{figure}

\begin{figure}
  \includegraphics[width=3.5in]{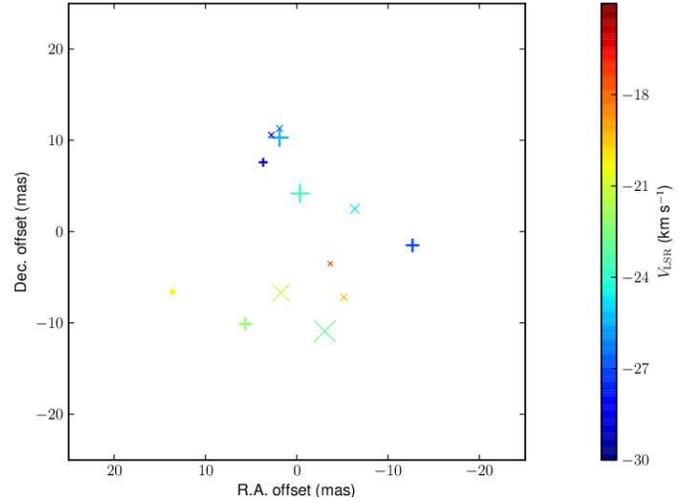}
  \caption{Color coded velocities of SiO masers surrounding R Aqr with ``+'' marking
    the locations of $\nu$=1,J=1-0 SiO masers and ``x'' for
    $\nu$=2,J=1-0 masers.
    Velocities are indicated by the color with a scale bar to the right.
    The size of each symbol is proportional to the $\sqrt{\quad}$ of the
    peak flux density.} 
  \label{FigR_Aqr_Vel}%
\end{figure}

The locations of the masers are plotted on the gray scale image of the 
continuum in Figure \ref{FigR_Aqr}.
The velocities and relative strengths of the masers are shown in
Figure \ref{FigR_Aqr_Vel}.
As seen from Figure \ref{FigR_AqrSpec}, the SNR of the masers exceed
100 so the uncertainty of the maser locations due to image noise is
less than 0.3 mas.

The off--source RMS in the continuum image is 36 $\mu$Jy beam$^{-1}$
and the CLEAN restoring beam is 50 $\times$ 36 mas at position angle
-1$^\circ$. 
The continuum component in which the masers are imbedded has a peak
flux density of 6.1 $\pm$ 1.2
mJy and an integrated flux density of 10.8 $\pm$ 2.2
mJy and a deconvolved Gaussian size of 58.9 $\pm$ 1.6 mas $\times$
16.9 $\pm$ 1.8 mas at a position angle of 17$^\circ$ $\pm$ 1$^\circ$.
The Gaussian brightness temperature of this component is 6,550 $\pm$
1490 K. 
For the equivalent uniform disk, the brightness temperature is 3,690
 $\pm$ 840 K.
The integrated flux density of the emission visible in Figure
\ref{FigR_Aqr} is 40 mJy.
No continuum or masers were detected outside of the region shown in
Figure \ref{FigR_Aqr}.
\section{Discussion}
\subsection{Mira}
The continuum emission from Mira A is slightly elongated 40$^\circ$
from the direction of Mira B. 
Otherwise there is no hint in the present results of an interaction.
While not well defined, the maser ring is basically round.
Numerous measurements of the 7 mm SiO masers are available in the
literature. 
The maser rings seen at much higher resolution in \cite{Cotton2004}
and \cite{Cotton2006} are even less well defined but also show little
hint of an interaction.
While there are variations in maser velocity around the ring seen in
Figure \ref{FigMira_Vel}, this is commonly seen in AGB maser rings and
shows no obvious effect of the companion. 
SiO masers tracing the interaction of the AGB star with a close
companion was reported in  $\pi^1$ Gru \citep{Homan2020}
although with a closer companion (6 AU) than Mira B is to A. 

\cite{RM07} give a uniform disk size for Mira A of 54 $\pm$ 4 mas
$\times$ 50 $\pm$ 4 mas at position angle -30$^\circ$ $\pm$ 50 with a
flux density of 4.8 $\pm$ 0.2 mJy and a brightness temperature of 1680 
$\pm$ 250 K in 2000 October.
This is somewhat smaller than our uniform disk equivalent of
63.0$\times$52.6 mas. 
The \cite{RM07} flux density of 4.8 mJy is in good agreement with our
value of 5.0 mJy. 
Our larger size results in a lower estimated uniform disk brightness
temperature, 1300 $\pm$ 270 K at 45 GHz; however,
this difference is within the errors.
The alignment of the SiO  maser ring with Mira A in \cite{RM07}
generally agrees with that in Figure \ref{FigMira}; centered on the
photosphere. 
They do not detect Mira B.

Observations of Mira A in 2014 February by \cite{Matthews2015} give an
elliptical Gaussian fit at 46 GHz of 37.5 $\pm$ 2.1 $\times$ 31.7
$\pm$ 2.0 mas at position angle 147$^\circ$ $\pm$ 6$^\circ$ in excellent
agreement with our 2019 August results.
However, the flux density quoted by \cite{Matthews2015}, 7.6 $\pm$ 1.5
mJy, is 1.5 $\sigma$ higher than our 5.0 mJy and their uniform
disk brightness temperature of 2110 $\pm$ 440 K is correspondingly
higher than our 1300 $\pm$ 270 K. 
The difference between the current results and those of
\cite{Matthews2015} is 1.5 $\sigma$.
\cite{Matthews2018} presents evidence for variations in the sizes and
shapes of several other Mira variables so the differences between our
results and those of \cite{RM07} and \cite{Matthews2015} may simply
reflect changes in Mira A.

The observations of \cite{Matthews2015} detected Mira B and give a
deconvolved Gaussian size at 46 GHz of 18 $\pm$ 5 $\times$  1 $\pm$ 6
mas at position angle 172$^\circ$ $\pm$ 30$^\circ$ with a flux density of
0.97 $\pm$ 0.2 mJy.
This is in good agreement with our upper limit of 20 mas
although the star is at most marginally resolved in both cases.
Since Mira B is thought to be a white dwarf, any resolution of this
object is very unlikely to be of the star itself but more likely an
accretion disk around it.

While we detect no emission away from Mira A and B, \cite{Wong2016}
show ALMA observations of a plume of molecular gas in an arc from
Mira A up to 3'' in the direction away from Mira B but do not
associate it with an effect by Mira B. 
They also report that the SiO density drops sharply past 4 stellar radii.

ALMA observations by \cite{Kaminski2016} of gas phase AlO, the
precursor molecule to Al$_2$O$_3$ in aluminum dust, shows that the
density peaks at about 2R$\star$. 
This is near the size of the radio photosphere and the SiO maser
ring seen in Figure \ref{FigMira}.

\newpage
\subsection{R Aquarii}
\begin{figure}
  \includegraphics[width=3.0in]{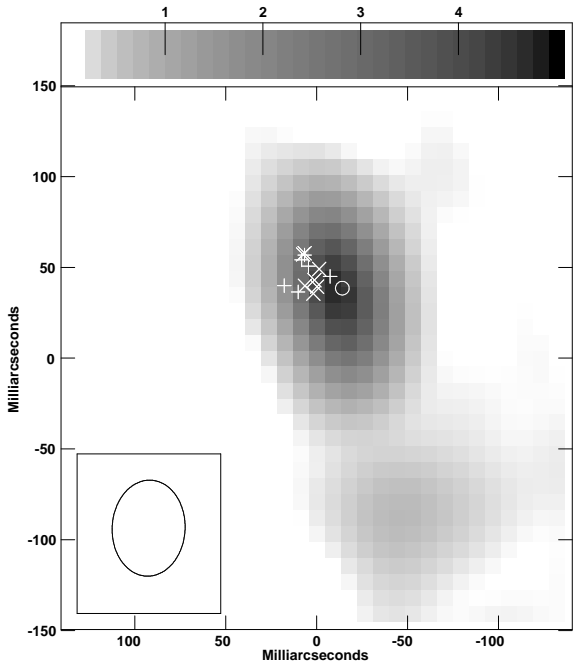}
  \caption{Like Figure \ref{FigR_Aqr} but a closeup of the emission
    enclosing the stars and a scalebar labeled in mJy beam$^{-1}$
The circle shows the location of the secondary as estimated from the
orbit of \cite{Bujarrabal2018}.}
  \label{FigR_Aqr_Close}%
\end{figure}

The orbit of the secondary star in R Aqr has been the subject of
debate for a number of years.
\cite{Hollis1997b} suggest that an extended feature seen in a VLA 7 mm
image corresponded to the secondary.
However, \cite{Schmid2017} dispute this interpretation and on the
basis of high spatial resolution H$_\alpha$ imaging suggest another.
A spectroscopic orbit has been determined by
\cite{Gromadzki2009} who derive an eccentric orbit ($e$=0.25) with a
43.6 year period but cannot determine where on the sky the secondary
appears. 
Detection of the secondary star by \cite{Bujarrabal2018} allowed them
to update the orbit of \cite{Gromadzki2009} giving an estimate of
the location at the epoch of the observations presented here.
The region of emission enclosing the stars is shown in Figure
\ref{FigR_Aqr_Close} with the estimated location of the white dwarf
shown as a circle.
As for Mira, Figure \ref{FigR_Aqr_Vel} shows no clear signature of an
effect of the secondary on the velocity structure of the ring.

However, ALMA observations by \cite{Bujarrabal2018} show plumes of CO
gas which they intreprete as arising from the outflow from the AGB star
being focused into its orbital plane by the white dwarf companion
maximizing the accretion onto the secondary.
They also detect continuum emission between the stars with a
spectral index suggesting dust.
Such structure would be obscured at 7 mm wavelength in 2019 by the
inner jet emission seen in Figure \ref{FigR_Aqr_Close}.

The continuum emission seen in the ALMA 0.8 mm observations are
not expected  to be the same as the 7 mm continuum emission as the
former is dominated by the photosphere and the dust forming around
the star and the latter is from free-free emission from ionized 
material in the circumstellar envelop.

This system contains a very extended and complex jet to the
north--east and south--west which is presumed to arise from the
secondary star \citep{Hollis1997a,Hollis1997b,Schmid2017}.
The jet material is quite visible in the H$_\alpha$ images presented
in \cite{Schmid2017} which also include the potential secondary.
The 7 mm image in Figure \ref{FigR_Aqr} is strikingly similar to
several of the H$_\alpha$ features south-west of the AGB star shown in
\cite{Schmid2017}, especially their features C, A$_{\rm SW}$, 
B$_{\rm SW}$ and C$_{\rm SW}$ and are approximately indicated in Figure
\ref{FigR_Aqr}. 
These features in Figure \ref{FigR_Aqr}, with the partial exception of
C, are presumably related to the jet.
This correspondance is puzzling given the 5 years between the
observations and the very short recombination times of these features
estimated by \cite{Schmid2017}.
This suggests either longer recombination times or a continuing source
of ionization.

The continuum feature associated with the SiO masers in Figures
\ref{FigR_Aqr} and \ref{FigR_Aqr_Close} is elongated in the general
direction of the jet and is not centered on the maser ring but rather
on the companion white dwarf. 
This suggests that this feature also encloses the secondary and some
of its emission is from the jet.
The Gaussian brightness temperature of this feature, 6,550  $\pm$ 970 K
is greatly in excess of that expected for the photosphere of an AGB
star further suggesting that much of the emission is from the
accretion disk/jet which is nearly aligned on the sky with the AGB star.


The circumstellar masers seen in Figures \ref{FigR_Aqr} and
\ref{FigR_Aqr_Close} show little evidence of being affected by the jet
or the secondary star. 
The VLBI image of the masers in R Aqr shown in \cite{Ragland2008}
show masers streaming off in the direction of the north--east jet in
2006 September but those in \cite{Cotton2004,Cotton2006} do not.

  The continuum emission see in Figures \ref{FigR_Aqr} and
\ref{FigR_Aqr_Close} is similar in shape and orientation to that
reported by \cite{Hollis1997b} from 1996, a half orbital period earlier,
but with a different offset from the SiO masers marking the atmosphere
of the AGB star.
This suggests that in both cases, the emission is dominated by the jet
from the accretion disk around the white dwarf and that the image of
\cite{Hollis1997b} does show the location of the secondary.

Unfortunately, the poor absolute astrometry resulting from the
atmospheric instability in the current data prevent them from being very
useful quantitatively in refining the orbit.

The distributions of masers in AGB atmospheres are extremely
variable as they track conditions suitable for maser amplification
rather than following concentrations of material.  
Observations several months apart, as shown in the references for VLBI
observations usually show very different distributions.  
The frequent monitoring of the Mira TX Cam \citep{Gonidakis2013}
shows very turbulent behavior. 

The nature of the SiO maser distribution in January 2001 from
\cite{Cotton2004} was of a very different character than at much later
epochs in showing a high degree of symmetry across the ring.  
In April 2001 the general velocity pattern of the masers was
consistent with a general rotation but without much evidence for an
equatorial disk. 
\cite{Hollis2001} claim some evidence for differential rotation in Dec
2000 as well as earlier with a pole of the rotation  at a position
angle of 150$^\circ$.
By 2006 \citep{Ragland2008} the velocity structure was incompatible
with the differential rotation of 2000-2001 and the 2019 velocity
structure in Figure \ref{FigR_Aqr_Vel} is also incompatible.  
The 2000-2001 event may have been a transient feature although the
orientation of the apparent pole is difficult to explain by an
interaction with either the gravity of, or the jet coming from, the
secondary. 
The jet is closer to the direction of the pole of the putative
rotation rather than orthogonal to it and the orbit of
\cite{Bujarrabal2018} shows a substantial separation between the two
stars in 2001.

\section{Conclusions}
Observations of 7 mm continuum and SiO masers in three symbotic AGB
stars are reported. 
The continuum data were phase referenced to the masers allowing
sensitive imaging and accurate registration.
Techniques for photometric calibration and this phase referencing were
developed to exploit these data.

No masers were detected in W Aquilae which prevented a sensitive
continuum image from being made.
Both Mira and R Aquarii have strong masers which allows deep
continuum imaging.
The image of Mira shows the masers well centered on the photosphere of
Mira A and the white dwarf Mira B was well detected.
No clear evidence of an interaction between the white dwarf and the
envelope of Mira A was found.

The inner portion of the jet in R Aquarii is visible in the continuum
image and the radio emission around the AGB star is extended in the
direction of the jet.
The continuum emission is centered on the location of the white dwarf
estimated from the orbit of \cite{Bujarrabal2018} rather than on the
AGB star as indicated by the SiO masers.
That, plus the high brightness temperature of this feature, indicates
that it is dominated by emission from the accretion disk + jets
associated with the white dwarf.
\cite{Bujarrabal2018} also shows gas apparently flowing from the AGB star
onto the secondary.
The alignment of the continuum peak of R Aqr with the predicted
location from the orbit of the secondary by \cite{Bujarrabal2018}
appears to support their orbit.

\begin{acknowledgements}
We thank B. Pimpanuwat/ATOMIUM consortium for the J=5-4 and J=6-5
measurements.
This work made use of the AAVSO International Database contributed by
observers worldwide. 

\end{acknowledgements}
\vspace{5mm}

\bibliography{Miras2019}{}
\bibliographystyle{aasjournal}

\end{document}